%% file: arma_graph_filter.tex
\documentclass{article}
\usepackage{spconf}
\include{use_packages}
\include{my_def_preamble}
\include{def_abr}
\def\BibTeX{{\rm B\kern-.05em{\sc i\kern-.025em b}\kern-.08em
    T\kern-.1667em\lower.7ex\hbox{E}\kern-.125emX}}

\begin{document}
\title{WLS Design of ARMA Graph Filters Using Iterative Second-Order Cone Programming}

\name{Darukeesan~Pakiyarajah$^{\star}$, Chamira~U.~S.~Edussooriya$^{\dagger,\ddagger}$\vspace{-5mm}} 
\address{$^{\star}$Dept. of Electrical and Electronic Engineering, University of Jaffna, Sri Lanka \\ $^{\dagger}$Dept. of Electronic and Telecommunication Engineering, University of Moratuwa, Sri Lanka \\ $^{\ddagger}$Dept. of Electrical and Computer Engineering, Florida International University, FL, USA \\
{\small Emails: darukeesan@eng.jfn.ac.lk, chamira@uom.lk}}

%

%


\maketitle

\begin{abstract}
We propose a \gls{wls} method to design \gls{arma} graph filters. We first express the \gls{wls} design problem as a numerically-stable optimization problem using Chebyshev polynomial bases. We then formulate the optimization problem with a nonconvex objective function and linear constraints for stability. We employ a relaxation technique and convert the nonconvex optimization problem into an iterative second-order cone programming problem. Experimental results confirm that \gls{arma} graph filters designed using the proposed \gls{wls} method have significantly improved frequency responses compared to those designed using previously proposed \gls{wls} design methods.  
\end{abstract}
%
%
\glsresetall
\section{Introduction}
\label{sec:int}
Graph signals can represent data associated with irregular structures, for example, in social networks, sensor networks, power grids and transportation networks~\cite{Shu2013a,San2013,Ort2018}. The introduction of the graph Fourier transform~\cite{San2013} has led to the processing of graph signals in the graph frequency domain, where graph filters constitute a key class of graph systems. A graph filter shapes the spectral content of a graph signal and are employed in applications such as signal denoising and smoothing~\cite{zhang18, isufi17}, classification~\cite{ma16}, clustering~\cite{tremblay16}, analog network coding~\cite{Sag2017a}, and signal reconstruction~\cite{isufi18}.

Graph filters are realized as \gls{fir} filters with polynomial frequency responses, and \gls{arma} filters with a rational frequency responses and \glspl{iir}. In the case of the time-varying graph signals, \gls{arma} filters have the ability to filter graph signals in both the graph frequency domain and regular temporal frequency domain~\cite{isufi17b}. Several methods have been proposed to design \gls{arma} graph filters including rational Butterworth filter~\cite{shi15}, rational Chebyshev filter~\cite{Rim20}, spectral-transformation-based methods~\cite{tseng20c}, and optimization methods~\cite{Aittomaki19, Liu19}. 

The optimization methods, unlike other methods, lead to graph filters having frequency responses optimal in either \gls{wls} or minimax sense. There are two fundamental challenges associated with the optimization methods for \gls{arma} graph filters: the optimization problem is nonconvex and highly susceptible to numerical instability. The \gls{wls} method proposed in~\cite{Aittomaki19} convert the nonconvex optimization problem to a convex optimization problem by modifying the objective function. Therefore, \gls{arma} graph filters designed using this method are not necessarily optimal in the \gls{wls} sense~\cite{Lu98}. In~\cite{Liu19}, an iterative scheme is employed without any modification to the nonconvex objective function. However, this \gls{wls} method does not incorporate stability constraints, possibly leading to unstable designs, and does not monotonically converge for some designs.

In this paper, we propose a \gls{wls} method to design \gls{arma} graph filters with \emph{a nonconvex objective function and linear constraints for stability}. We achieve our formulation by converting the filter design problem into a numerically-stable optimization problem using Chebyshev polynomial bases and employing a relaxation technique. We employ an iterative \gls{socp} scheme to solve the optimization problem. Experimental results confirm that \gls{arma} graph filter designed using the proposed \gls{wls} method has \emph{significantly improved} frequency response compared to those designed using previously proposed \gls{wls} design methods, in particular, $88\%$ reduction in the passband ripple and $17$ dB reduction in the \gls{sse} are achieved compared to the state-of-the-art \gls{wls} method~\cite{Aittomaki19}.

\section{Proposed WLS Design Method}
\label{sec:method}
\subsection{Problem Formulation}
\label{sec:probform}
The frequency response $h(\lambda)$ of an \gls{arma} graph filter of order $(P,Q)$ can be expressed as~\cite{isufi17b}
\begin{align}
h(\lambda) = \frac{\sum\limits_{p=0}^{P} b_p \lambda^p}{1+\sum\limits_{q=1}^{Q} a_q \lambda^q},
\label{eq:freqres}
\end{align}
where $\lambda$ is the graph spectral frequency. In the design of the \gls{arma} graph filter, we need to determine the coefficients $b_0, b_1, \ldots, b_P$, and $a_1, a_2, \ldots, a_Q$ such that the spectral response $h(\lambda)$ approximates a given ideal spectral response $h_d(\lambda)$ as closely as possible. Because the denominator and numerator polynomials of $h(\lambda)$ are linear combination of monomial basis $\{1, \lambda, \lambda^2, \cdots, \lambda^M\}$ and $\lambda \in [0,2]$, the conventional optimization techniques employed in digital \gls{iir} filter designs are highly susceptible to numerical instability for the designs of higher-order \gls{arma} filters of the form given in~\eqref{eq:freqres}. Due to the limited space, we do not present definitions of graph signals and graph filters, and the reader is refereed to~\cite{San2013} and \cite{Ort2018}.

To convert the filter design problem into a numerically stable problem, without loss of generality, we express $h(\lambda)$ in the form
\vspace{-2ex
}
\begin{align}
h(\lambda) = \frac{\sum\limits_{p=0}^{P} \beta_p T_p(1-\lambda)}{1+\sum\limits_{q=1}^{Q} \alpha_q T_q(1-\lambda)},
\label{eq:freqres2}
\end{align}
where $T_n(x)$ is the Chebyshev polynomial of order $n$, which satisfies the recursive relation $T_n(x)=2xT_{n-1}(x)-T_{n-2}(x)$ for $n \geq 2$ with $T_0(x)=1$ and $T_1(x)=x$. We note that the use of Chebyshev polynomials for the design of \gls{iir} graph filters is inspired from previous works on the Chebyshev polynomial based designs of \gls{fir} graph filters~\cite{Shuman18, tseng21}. Now, we express $h(\lambda)$ as $h(\lambda) = \frac{\bs{c}_P\tr(\lambda)\bs{\beta}}{1+\bs{c}_Q\tr(\lambda)\bs{\alpha}}$,
where $\bs{\alpha}=[\alpha_1, \alpha_2, \hdots,\alpha_Q]\tr$, $\bs{\beta}=[\beta_0, \beta_1, \hdots,\beta_P]\tr$, $\bs{c}_P(\lambda)=[T_0(1-\lambda), T_1(1-\lambda), \hdots,T_P(1-\lambda)]\tr$, and $\bs{c}_Q(\lambda)=[T_1(1-\lambda), T_2(1-\lambda), \hdots,T_Q(1-\lambda)]\tr$. Here, $\bs{c}_p$ and $\bs{c}_Q$ are the orthogonal shifted Chebyshev polynomial bases. The \gls{wls} design of the \gls{arma} graph filter can then be expressed as the minimization problem given by 
\begin{subequations}
\begin{align}
      \underset{\bs{\alpha}, \bs{\beta}}{\text{minimize}} \:\: & \quad J(\bs{\alpha}, \bs{\beta}) \label{eq:objfunc}\\ 
      \text{subject to:} & \quad h(\lambda) \text{ is stable for } \lambda \in [0,2]. \label{eq:stabcons}
\end{align}
\end{subequations}
The objective function $J(\bs{\alpha}, \bs{\beta})$ is defined by  $J(\bs{\alpha}, \bs{\beta})=\int_0^2 W(\lambda) \left[h(\lambda) - h_d(\lambda)\right]^2 \df \lambda$, 
where $W(\lambda)$ is a nonnegative weighting function. 
Note that the properties $|1-\lambda| \leq 1$ and $|T_n(x)|\leq1$ for $|x| \leq 1$ of the selected orthogonal Chebyshev polynomial bases \emph{significantly improve} the numerical stability of the optimization problem in~\eqref{eq:objfunc} and~\eqref{eq:stabcons} compared to an optimization problem corresponding to the monomial basis~\cite[ch. 7.3.4]{Heath96ScientificCA}. Thus, we can employ the conventional optimization techniques to design the \gls{arma} filters of the form given in~\eqref{eq:freqres2} for reasonably large values of $P$ and $Q$.

\subsection{Stability Condition}
\label{sec:stabcond}
A causal and stable \gls{arma} filter can be obtained when $h(\lambda)$ does not have any pole in the range $\lambda\in [0,2]$~\cite{Liu19}, or equivalently, when $1+\bs{c}_Q \tr (\lambda)\bs{\alpha}\neq 0$ for $\lambda \in [0,2]$. A practical way to incorporate this condition in optimization problems is by slightly modifying this constraint as~\cite{Lu98}
\begin{align}
      1+\bs{c}_Q \tr (\lambda) \bs{\alpha}\geq \epsilon \quad \text{for } \lambda \in [0, 2],
\label{eq:stabcons2}
\end{align}
where $\epsilon$ is a small positive quantity. Now, we can express the constraint in \eqref{eq:stabcons2} equivalently as a linear constraint given by
\begin{align}
      -\bs{c}_Q \tr (\lambda) \bs{\alpha}\leq 1-\epsilon \quad \text{for } \lambda \in [0, 2].
\label{eq:stabcons3}
\end{align}

\subsection{Iterative Scheme}
\label{sec:iterscheme}
In this subsection, we present the iterative scheme required to solve the optimization problem in~\eqref{eq:objfunc} and~\eqref{eq:stabcons}. Using~\eqref{eq:freqres2}, we express $J(\bs{\alpha},\bs{\beta})$ as
\begin{align}
      J(\bs{\alpha}, \bs{\beta}) & =\int_0^2 W(\lambda) \left[\frac{\bs{c}_P\tr(\lambda)\bs{\beta}}{1+\bs{c}_Q\tr(\lambda)\bs{\alpha}} - h_d(\lambda)\right]^2 \df \lambda \notag \\
      & =\int_0^2 \frac{W(\lambda)}{({1+\bs{c}_Q\tr(\lambda)\bs{\alpha}})^2} \Big[\bs{c}_P\tr(\lambda)\bs{\beta} \notag \\ 
      & \qquad \qquad \qquad - h_d(\lambda)\bs{c}_Q\tr(\lambda)\bs{\alpha} - h_d(\lambda)\Big]^2 \df \lambda,
\label{eq:objfunc3}
\end{align}
which is a \emph{nonconvex} function. We note that the \gls{wls} method proposed in~\cite{Aittomaki19} neglects the term 
$({1+\bs{c}_Q\tr(\lambda)\bs{\alpha}})^2$ beneath $W(\lambda)$ and minimizes the so-called modified error given by
\begin{align}
      \Hat{J}(\bs{\alpha}, \bs{\beta}) = \int_0^2 W(\lambda) \Big[\bs{c}_P\tr(\lambda)\bs{\beta} - h_d(\lambda)\bs{c}_Q\tr(\lambda)\bs{\alpha} \notag \\ \qquad \qquad- h_d(\lambda)\Big]^2 \df \lambda. 
\label{eq:modierr}
\end{align}
In this case, $\bs{\alpha}$ and $\bs{\beta}$ that minimize $\Hat{J}(\bs{\alpha}, \bs{\beta})$ do not necessarily minimize~\eqref{eq:objfunc3}. In general, the solution obtained by minimizing~\eqref{eq:modierr} is not optimal in the
\gls{wls} sense~\cite{Lu98}. To incorporate the term $({1+\bs{c}_Q\tr(\lambda)\bs{\alpha}})^2$ into the optimization problem, we adopt an iterative scheme given by 
\begin{align}
      J(\bs{\alpha}_k, \bs{\beta}_k) & = \int_0^2 W_k(\lambda) \Big[\bs{c}_P\tr(\lambda)\bs{\beta}_k - \notag \\ & \qquad \qquad h_d(\lambda)\bs{c}_Q\tr(\lambda)\bs{\alpha}_k - h_d(\lambda)\Big]^2 \df \lambda,  \label{eq:iterscheme} \\
      W_k(\lambda) &=\frac{W(\lambda)}{({1+\bs{c}_Q\tr(\lambda)\bs{\alpha}_{k-1}})^2},   
\label{eq:iterweighfunc}
\end{align}
where $k=0,1,2,\hdots$, and $\bs{\alpha}_k$, $\bs{\beta}_k$ are variables to be determined in the $k^{\text{th}}$ iteration. The initial value $\bs{\alpha}_0$ required to compute $W_1(\lambda)$ should be taken such that it satisfies the condition in~\eqref{eq:stabcons3}. For instance, $\bs{\alpha}_0 = [0, 0, \hdots, 0]\tr$ is a suitable initial value. 

The nonconvex objective function $J(\bs{\alpha}, \bs{\beta})$ in~\eqref{eq:objfunc3} generally has multiple minima. Therefore, the solution obtained by iterative scheme in~\eqref{eq:iterscheme} and~\eqref{eq:iterweighfunc} can be expected to be only a local minimum of $J(\bs{\alpha}, \bs{\beta})$~\cite{Lu98}. The systematic procedure required to solve the iterative scheme in~\eqref{eq:iterscheme} and~\eqref{eq:iterweighfunc}, subject to the stability constraint in~\eqref{eq:stabcons3}, is presented in Sec.~\ref{sec:iteralgo}.

\subsection{SOCP Approach}
\label{sec:socpprblm}
In this subsection, we convert the optimization problem of minimizing $J(\bs{\alpha}_k, \bs{\beta}_k)$ subject to~\eqref{eq:stabcons3} into an \gls{socp} problem. To this end, we express $J(\bs{\alpha}_k, \bs{\beta}_k)$ as,
\begin{align}
      J(\bs{h}_k) & = \int_0^2 W_k(\lambda) \Big[\bs{d}(\lambda)\tr\bs{h}_k - h_d(\lambda)\Big]^2 \df \lambda, 
\label{eq:objfunc4}
\end{align}
where $\bs{d}(\lambda)= \begin{bmatrix}\bs{c}_P\tr(\lambda) & -h_d(\lambda)\bs{c}_Q\tr(\lambda)\end{bmatrix}\tr$, and \sloppy $\bs{h}_k = \begin{bmatrix}\bs{\beta}_k\tr & \bs{\alpha}_k\tr\end{bmatrix}\tr$. The expression in~\eqref{eq:objfunc4} can be expanded as 
\begin{align}
      J(\bs{h}_k) & = \bs{h}_k\tr\bs{Q}_k\bs{h}_k-2\bs{q}_k\tr\bs{h}_k+\const{p}_k, 
\label{eq:objfunc5}
\end{align}
where $\bs{Q}_k = \int_0^2 W_k(\lambda) \bs{d}(\lambda) \bs{d}\tr(\lambda) \: \: \df \lambda$, $\bs{q}_k = \int_0^2 W_k(\lambda) h_{d}(\lambda) \bs{d}(\lambda) \: \:  \df \lambda$, and $\const{p}_k = \int_0^2 W_k(\lambda) h_d^2(\lambda) \: \: \df \lambda$.
We then express~\eqref{eq:objfunc5} as 
\begin{align}
      J(\bs{h}_k) & = \bs{h}_k\tr \bs{Q}_k^{\mathrm{T}/2}\bs{Q}_k^{1/2}\bs{h}_k-2\bs{h}_k\tr\bs{Q}_k^{\mathrm{T}/2}\bs{Q}_k^{-\mathrm{T}/2}\bs{q}_k+ \const{p}_k \notag \\
    &= \|\widehat{\bs{Q}}_k\bs{h}_k-\widehat{\bs{q}}_k\|_2^{2}-\widehat{\bs{q}}_k\tr\widehat{\bs{q}}_k+\const{p}_k, 
\label{eq:objfunc6}
\end{align}
where $\widehat{\bs{Q}}_k=\bs{Q}_k^{1/2}$, $ \widehat{\bs{q}}_k= \bs{Q}_k^{-\mathrm{T}/2}\bs{q}_k$, and $\|\cdot\|_2$ is the $2$-norm of a vector. Because $\const{p} - \widehat{\bs{q}}_k\tr\widehat{\bs{q}}_k$ is a constant term for a given $k$, we can equivalently formulate the optimization problem of minimizing $J(\bs{h}_k)$ subject to~\eqref{eq:stabcons3} as
\begin{subequations} \label{eq:wlsoptprob}
\begin{align}
\underset{\bs{h}_k}{\text{minimize}} \:\: & \quad \eta_k 
\label{eq:wlsobjfunc} \\
\text{subject to:} & \quad \|\widehat{\bs{Q}}_k\boldsymbol{h}_k-\widehat{\bs{q}}_k\|_2 \leq \eta_k,\label{eq:wlsconstr1} \\
& \quad \bs{g}\tr (\lambda) \bs{h}_k\leq 1-\epsilon \quad \text{for } \lambda \in [0, 2], \label{eq:wlsconstr2}
\end{align}
\end{subequations}
where $\eta_k$ is an upper bound on $\|\widehat{\bs{Q}}_k\boldsymbol{h}_k-\widehat{\bs{q}}_k\|_2$, $\bs{g}(\lambda) = \begin{bmatrix} \bs{O}_{1 \times (P+1)} & -\bs{c}_Q\tr(\lambda)\end{bmatrix}\tr$, and $\bs{O}_{M\times N}$ is an $M \times N $ zero matrix.

Next, we consider the discretized version of the optimization problem in~\eqref{eq:wlsoptprob} with a dense set of values for $\lambda$, i.e., $\Lambda = \{\lambda_{d,0}, \lambda_{d,1},\lambda_{d,2},\cdots,\lambda_{d,L}\} \subseteq [0,2]$. The straightforward numerical approximations for $\boldsymbol{Q}_k$ and $\boldsymbol{q}_k$ are 
\begin{align}
    & \bs{Q}_k\approx \sum_{j=0}^LW_k(\lambda_{d,j}) \bs{d}(\lambda_{d,j}) \bs{d}\tr(\lambda_{d,j})\\
    & \bs{q}_k\approx \sum_{j=0}^L W_k(\lambda_{d,j}) h_{d}(\lambda_{d,j}) \bs{d}(\lambda_{d,j}),
    \label{eq:dicapprox}
\end{align}
respectively~\cite{Lu2011a, Pak2021}. Then we can express the optimization problem in~\eqref{eq:wlsoptprob} as an \gls{socp} problem, in the standard form, as~\cite[ch. 14.7]{antoniou2007optimization}
\begin{subequations} \label{eq:wlsoptprob2}
\begin{align}
\underset{\bs{x}_k}{\text{minimize}} \:\: & \quad \bs{f}\tr\bs{x}_k 
\label{eq:wlsobjfunc2} \\
\text{subject to:} & \quad \|\bs{A}_k\tr\boldsymbol{x}_k+\bs{b}_k\|_2 \leq \bs{f}\tr\bs{x}_k,\label{eq:wlsconstr1_2} \\
& \quad \bs{B}\tr \bs{x}_k\leq (1-\epsilon) \bs{e}_{L+1}, \label{eq:wlsconstr2_2}
\end{align}
\end{subequations}
where $\bs{f} = \begin{bmatrix} 1 & 0 & 0 & \cdots & 0 \end{bmatrix}\tr_{1\times (P+Q+2)}$, $\bs{x}_k = \begin{bmatrix} \eta_k & \bs{h}_k\tr \end{bmatrix}\tr$, $\bs{A}_k = \begin{bmatrix} \bs{O}_{(P+Q+1)\times 1} & \bs{\widehat{\bs{Q}}_k} \end{bmatrix}\tr$, $\bs{b}_k = -\widehat{\bs{q}}_k$, $\bs{B}=\begin{bmatrix} \bs{O}_{(L+1)\times 1} & \begin{bmatrix} \bs{g}(\lambda_{d,0}) & \bs{g}(\lambda_{d,1}) & \hdots & \bs{g}(\lambda_{d,L}) \end{bmatrix}\tr \end{bmatrix}\tr$, and  $\bs{e}_{L+1} = \begin{bmatrix} 1 & 1 & \cdots & 1 \end{bmatrix}\tr_{1\times (L+1)}$. In MATLAB, the \gls{socp} problem in \eqref{eq:wlsoptprob2} can be effectively solved using convex optimization toolboxes such as CVX~\cite{cvx, gb08}. 


\subsection{Convergence of the Iterative Scheme}
\label{sec:convergence}
We now present a modification required to ensure the convergence of the iterative scheme proposed in Sec.~\ref{sec:iterscheme}. Because the original optimization problem in~\eqref{eq:objfunc} and~\eqref{eq:stabcons} is nonconvex, the proposed iterative technique is not necessarily guaranteed to converge for all filter parameters. To handle this issue, we adopt the relaxation technique proposed in~\cite{Lu98}, and slightly modify the iterative scheme. Let $\Phi$ be the operator that maps the initial point $\bs{x}_{k-1}$ to the solution of the \gls{socp} problem in~\eqref{eq:wlsoptprob2}. Then, the iterative scheme proposed in Sec.~\ref{sec:iterscheme} can be expressed as $\bs{x}_{k} = \Phi(\bs{x}_{k-1})$.
Now, we modify this to 
\begin{align}
    \bs{x}_{k} = \gamma \Phi(\bs{x}_{k-1}) + (1-\gamma)\bs{x}_{k-1},
    \label{eq:modirelax}
\end{align}
where $\gamma \in [0, 1]$ is a relaxation constant. Equivalently, we can state that $\bs{x}_{k}$ is obtained by combining the solution of~\eqref{eq:wlsoptprob2} with the initial point $\bs{x}_{k-1}$ used in~\eqref{eq:wlsoptprob2}. There is no approach to theoretically determine appropriate $\gamma$ required for the convergence of the algorithm. However, we can experimentally obtain a feasible range of values for $\gamma$~\cite{Lu98}.

\subsection{Modified Iterative Algorithm}
\label{sec:iteralgo}
We now summarize the iterative algorithm described through Secs.~\ref{sec:probform} to \ref{sec:convergence} in this subsection, with a terminating conditions for the iterative scheme. The iteration begins by choosing an initial value for $\bs{x_0}$. At each iteration, matrices $\bs{A}_k$ and $\bs{b}_k$ are updated, the optimization problem in~\eqref{eq:wlsoptprob2} is solved, and $\bs{x}_k$ is updated according to~\eqref{eq:modirelax}. The iteration continues until $\|\bs{x}_k - \bs{x}_{k-1}\|_{\infty}$ is less than a prescribed tolerance $\delta_t$ or the number of iterations reach a predefined maximum iteration $k_{max}$. Here, $\|\cdot\|_{\infty}$ is the infinity-norm of a vector. At the end of this iterative process, the convergence is claimed, and $\bs{x}_k$ is considered as the optimal solution $\bs{x}_{opt}$ of the \gls{wls} optimization problem in~\eqref{eq:wlsoptprob2}. This modified algorithm is presented in Alg.~\ref{alg:the_alg}.

\begin{algorithm}[t!]
    \caption{Modified Iterative Algorithm} \label{alg:the_alg}
        \begin{algorithmic}[1]
            \Require $P,~Q, ~L, ~\Lambda, ~h_d(\lambda_{d,i}), ~W(\lambda_{d,i}), ~\epsilon, ~\gamma, ~\delta_t$, $k_{max}$
            \Init $k \gets 1$, ~$\bs{x}_0$
            \Def $\bs{f}, ~\bs{B}, ~\bs{e}_{L+1}$
            \Iter
            \State Compute $\bs{A}_k, ~\bs{b}_k$    \label{algstate:iterbegin}
            \State Solve the \gls{socp} problem in~\eqref{eq:wlsoptprob2} for $\bs{x}_k$
            \State $\bs{x}_k \gets \gamma\bs{x}_k + (1-\gamma)\bs{x}_{k-1}$
            \If{$\|\bs{x}_k-\bs{x}_{k-1}\|_\infty > \delta_t$ and $k < k_{max}$}
                \State $k \gets k+1$
                \State Go to step~\ref{algstate:iterbegin}
            \EndIf
            \Ensure $\bs{x}_{opt} \gets \bs{x}_k$
        \end{algorithmic}
\end{algorithm}
\vspace{-3ex}
\section{Experimental Results}
\label{sec:DesExmpls}
In this section, we present experimental results obtained for an example design of a lowpass \gls{arma} graph filter, and we compare the performance of the proposed method with the inverse solution and peak-error constrained \gls{wls} design in~\cite{Aittomaki19}, the iterative scheme in~\cite{Liu19}, and the least-square \gls{fir} filter in \cite{tseng20b} to confirm the effectiveness of the proposed method. We design all the filters using MATLAB and CVX~\cite{cvx, gb08} as the optimization toolbox.

We define the ideal frequency response of the filter as,
\begin{align}
h_d(\lambda) = 
\begin{cases} 
      1, & 0\leq \lambda \leq \lambda_p \\
      \text{don't care}, & \lambda_p < \lambda < \lambda_s \\
      0, & \lambda_s \leq \lambda \leq 2
\end{cases}
\label{eq:idealres}
\end{align}
where $\lambda_p$ and $\lambda_s$ are the passband and stopband edges of the lowpass filter, respectively. We select $\lambda_p = 0.5$ and $\lambda_s = 0.7$, respectively, which are the specifications employed in the design example considered in~\cite{Aittomaki19}. Furthermore, we select the weight function $W(\lambda)$ as $1$, $0$ and $1$ for $0\leq \lambda \leq \lambda_p$, $\lambda_p < \lambda < \lambda_s$, and $\lambda_s \leq \lambda \leq 2$, respectively.  
We select the order of the filter as $P=Q=11$ (the same as~\cite{Aittomaki19}) and the discretized values of $\lambda$ as $\lambda_{d,i}=\frac{2i}{L}$, where $L=500$. The total number of points in the passband and stopband is $401$. Furthermore, we select the parameters $\epsilon=10^{-5}$ (see~\eqref{eq:stabcons3}), $k_{max} = 25$, and $\delta_t = 2\times10^{-8}$ (see Alg.~\ref{alg:the_alg}). We use the initial value of $\bs{x}_0 = \bs{O}_{(P+Q+2)\times1}$ in this design. From experiments, we found that $\gamma \in [0.1, 0.5]$ ensures the convergence of the iterative algorithm, and we use $\gamma=0.25$ for this design. In this case, the algorithm converges after $7$ iterations.

\begin{figure}[t!]
\centering
\includegraphics[trim=0.2cm 0.2cm 0.2cm 0.75cm, clip=true, width=0.9\linewidth]{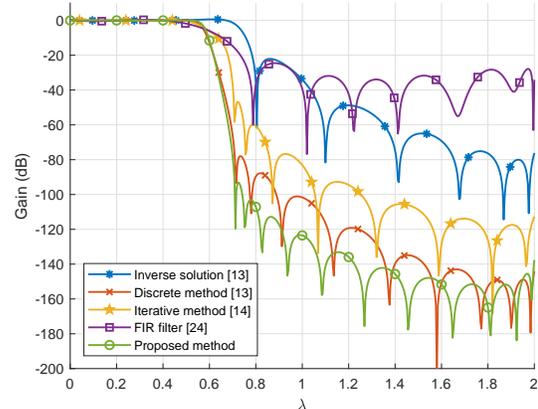}
\vspace{-2ex}
\caption{The magnitude response (in dB) of the lowpass graph filter designed using different WLS methods.} 
\label{fig:magres}
\vspace{-2ex}
\end{figure}

Next, we employ the previously proposed \gls{arma} graph filter design methods in~\cite{Aittomaki19}: discrete and inverse solution (see (25) in~\cite{Aittomaki19}), the iterative method in~\cite{Liu19}, and the \gls{wls} \gls{fir} graph filter design method in~\cite{tseng20b} to design graph filters for the same specifications. Here, we set the regularization parameter as $0.005$ for the inverse solution method, and the peak error constraint as $10^{-5}$ for the discrete method. Furthermore, we set norm penalty constant as $5.5\times10^{-6}$ for the \gls{fir} filter design~\cite{tseng20b}. The orders of the \gls{arma} graph filter is $11$ and that of the \gls{fir} graph filter is $22$. The iterative method in~\cite{Liu19} does not monotonically converge, and we consider the best filter design achieved in the $7^{\text{th}}$ iteration among the first $30$ iterations. We present the magnitude responses of the designed filters in Fig.~\ref{fig:magres}. We observe that the magnitude response of the \gls{arma} graph filter designed using the proposed method well approximates the desired magnitude response. Furthermore, we compare the maximum passband ripple ($\delta_p$), maximum stopband attenuation ($\delta_s$), and the \gls{sse} of the designed filters in Table~\ref{tab:comp}. It is evident that the proposed design method \emph{significantly outperforms} those proposed in~\cite{Aittomaki19}, \cite{Liu19} and \cite{tseng20b} with \emph{all the three parameters}. In particular, $\delta_p$ and \gls{sse} are reduced by $88\%$ and $17$ dB, respectively, comapred to~\cite{Aittomaki19}.

\begin{table}[t!]
\centering
\caption{Parameters of the magnitude responses achieved with the different designs of the lowpass graph filter.}
\vspace{-1ex}
\renewcommand{\arraystretch}{1.2}
\begin{tabular}{l r r r}
\hline
 \multirow{2}{*}{Method} & \multicolumn{3}{c}{Parameter} \\
  &  $\delta_p$ (dB) & $\delta_s$ (dB) & SSE (dB)\\
  \hline
  Inverse solution~\cite{Aittomaki19} & $0.1909$ & $1.5997$	& $7.5223$ \\
  Discrete~\cite{Aittomaki19}& $0.0134$	& $70.4361$	& $-50.5811$ \\
  Iterative~\cite{Liu19}& $0.2284$	& $43.7050$	& $-28.3991$ \\
  \gls{fir} filter~\cite{tseng20b} & $1.5305$ & $15.2755$	& $-1.1277$ \\
  Proposed & $\bs{0.0016}$	& $\bs{77.5931}$ & $\bs{-67.7455}$ \\
  \hline
\end{tabular}
\label{tab:comp}
\vspace{-3ex}
\end{table}

\section{Conclusion and Future Work}
\label{sec:conclns}
We propose a \gls{wls} design method for \gls{arma} graph filters by employing Chebyshev polynomials and iterative \gls{socp} scheme. In contrast to previously proposed \gls{wls} design methods, we consider the optimization problem without any modifications to the nonconvex objective function.  Experimental results confirm that \gls{arma} graph filters designed using the proposed \gls{wls} method have significantly improved frequency response compared to those designed with previously proposed \gls{wls} design methods. Future work includes the design of ARMA graph filters optimal in the minimax sense.

\section{Acknowledgment}
\label{sec:ack}
Authors thank the University of Moratuwa for financial support.

\bibliographystyle{IEEEtran}    
\bibliography{BibGraphFilDes}

\end{document}

%% file: use_packages.tex
\usepackage{cite}
\usepackage{url}
\usepackage{textcomp}

\usepackage{array}
\usepackage{multirow}
\usepackage{arydshln}

\usepackage{amsmath,amssymb,amsfonts}
\usepackage{stfloats}
\usepackage{nicefrac}

\usepackage{graphicx}
\usepackage{subfigure}
\usepackage{xcolor}
\usepackage{tikz,pgfplots}
\usetikzlibrary{arrows}

\usepackage[acronym]{glossaries}

\usepackage{algorithm}
\usepackage{algpseudocode}

\usepackage{caption}

\usepackage{amsmath}
\usepackage{mathtools}
\usepackage[capitalise,nameinlink]{cleveref}

\usepackage{enumitem}
\usepackage{array}


%% file: my_def_preamble.tex
\graphicspath{Figures/}
\newcommand{\const}[1]{\mathrm{#1}}		
\newcommand{\tr}{^\mathrm{T}}			
\newcommand{\df}{\mathrm{d}}			
\newcommand{\bs}[1]{\boldsymbol{#1}}	
\allowdisplaybreaks[4]					

\newcommand{\Init}{\item[\textbf{Initialize:}]}
\newcommand{\Def}{\item[\textbf{Define:}]}
\newcommand{\Iter}{\item[\textbf{Iteration:}]}

%% file: def_abr.tex
\newacronym{fir}{FIR}{finite-extent impulse response}
\newacronym{iir}{IIR}{infinite-extent impulse response}
\newacronym{psnr}{PSNR}{peak-signal-to-noise ratio}
\newacronym{snr}{SNR}{signal-to-noise ratio}
\newacronym{awgn}{AWGN}{additive white Gaussian noise}

\newacronym{nrmse}{NRMSE}{normalized-root-mean-square error}
\newacronym{cnn}{CNN}{convolutional neural network}
\newacronym{iid}{iid}{independent and identically distributed}
\newacronym{dft}{DFT}{discrete Fourier transform}
\newacronym{idft}{IDFT}{inverse discrete Fourier transform}
\newacronym{gft}{GFT}{graph Fourier transform}
\newacronym{igft}{IGFT}{inverse graph Fourier transform}
\newacronym{fft}{FFT}{fast Fourier transform}
\newacronym{fpga}{FPGA}{field-programmable gate array}

\newacronym{lp}{LP}{linear programming}
\newacronym{cpa}{CPA}{Chebyshev polynomial approximation}
\newacronym{ls}{LS}{least square}
\newacronym{arma}{ARMA}{autoregressive moving average}
\newacronym{wls}{WLS}{weighted least-square}
\newacronym{socp}{SOCP}{second-order cone programming}

\newacronym{sse}{SSE}{sum-of-square-error}

%% file: arma_graph_filter.bbl
\begin{thebibliography}{10}
\providecommand{\url}[1]{#1}
\csname url@samestyle\endcsname
\providecommand{\newblock}{\relax}
\providecommand{\bibinfo}[2]{#2}
\providecommand{\BIBentrySTDinterwordspacing}{\spaceskip=0pt\relax}
\providecommand{\BIBentryALTinterwordstretchfactor}{4}
\providecommand{\BIBentryALTinterwordspacing}{\spaceskip=\fontdimen2\font plus
\BIBentryALTinterwordstretchfactor\fontdimen3\font minus
  \fontdimen4\font\relax}
\providecommand{\BIBforeignlanguage}[2]{{%
\expandafter\ifx\csname l@#1\endcsname\relax
\typeout{** WARNING: IEEEtran.bst: No hyphenation pattern has been}%
\typeout{** loaded for the language `#1'. Using the pattern for}%
\typeout{** the default language instead.}%
\else
\language=\csname l@#1\endcsname
\fi
#2}}
\providecommand{\BIBdecl}{\relax}
\BIBdecl

\bibitem{Shu2013a}
D.~I. Shuman, S.~K. Narang, P.~Frossard, A.~Ortega, and P.~Vandergheynst, ``The
  emerging field of signal processing on graphs: Extending high-dimensional
  data analysis to networks and other irregular domains,'' \emph{IEEE Signal
  Process. Mag.}, vol.~30, no.~3, pp. 83--98, May 2013.

\bibitem{San2013}
A.~Sandryhaila and J.~M. Moura, ``Discrete signal processing on graphs,''
  \emph{IEEE Trans. Signal Process.}, vol.~61, no.~7, pp. 1644--1656, Apr.
  2013.

\bibitem{Ort2018}
A.~Ortega, P.~Frossard, J.~Kova{\v{c}}evi{\'c}, J.~M. Moura, and
  P.~Vandergheynst, ``Graph signal processing: Overview, challenges, and
  applications,'' \emph{Proc. IEEE}, vol. 106, no.~5, pp. 808--828, May 2018.

\bibitem{zhang18}
F.~Zhang and E.~Hancock, ``Graph spectral image smoothing using the heat
  kernel,'' \emph{Pattern Recognition}, vol.~41, pp. 3328--3342, Nov. 2008.

\bibitem{isufi17}
E.~Isufi and G.~Leus, ``Distributed sparsified graph filters for denoising and
  diffusion tasks,'' in \emph{Proc. IEEE Int. Conf. Acoust., Speech, Signal
  Process.}, 2017, pp. 5865--5869.

\bibitem{ma16}
J.~Ma, W.~Huang, S.~Segarra, and A.~Ribeiro, ``Diffusion filtering for graph
  signals and its use in recommendation systems,'' in \emph{Proc. IEEE Int.
  Conf. Acoust., Speech, Signal Process.}, Mar. 2016.

\bibitem{tremblay16}
N.~Tremblay, G.~Puy, R.~Gribonval, and P.~Vandergheynst, ``Compressive spectral
  clustering,'' in \emph{Proc. Int. Conf. Machine Learning}, vol.~48, 2016, pp.
  1002--1011.

\bibitem{Sag2017a}
S.~Segarra, A.~G. Marques, and A.~Ribeiro, ``Optimal graph-filter design and
  applications to distributed linear network operators,'' \emph{IEEE Trans.
  Sig. Process.}, vol.~65, no.~15, pp. 4117--4131, Aug. 2017.

\bibitem{isufi18}
E.~Isufi, P.~Di~Lorenzo, P.~Banelli, and G.~Leus, ``Distributed {Wiener}-based
  reconstruction of graph signals,'' in \emph{Proc. IEEE Stat. Sig. Process.
  Workshop}, 2018, pp. 21--25.

\bibitem{isufi17b}
E.~Isufi, A.~Loukas, A.~Simonetto, and G.~Leus, ``Autoregressive moving average
  graph filtering,'' \emph{IEEE Trans. Sig. Process.}, vol.~65, no.~2, pp.
  274--288, 2017.

\bibitem{shi15}
X.~Shi, H.~Feng, M.~Zhai, T.~Yang, and B.~Hu, ``Infinite impulse response graph
  filters in wireless sensor networks,'' \emph{IEEE Sig. Process. Letters},
  vol.~22, no.~8, pp. 1113--1117, 2015.

\bibitem{Rim20}
O.~Rimleanscaia and E.~Isufi, ``Rational {Chebyshev} graph filters,'' in
  \emph{2020 54th Asilomar Conf. Signals, Systems, Computers}, 2020, pp.
  736--740.

\bibitem{tseng20c}
C.-C. Tseng, ``Rational graph filter design using spectral transformation and
  {IIR} digital filter,'' in \emph{2020 IEEE Region 10 Conf. (TENCON)}, 2020,
  pp. 247--250.

\bibitem{Aittomaki19}
T.~Aittomaki and G.~Leus, ``Graph filter design using sum-of-squares
  representation,'' in \emph{Proc. European Signal Process. Conf.}, 2019, pp.
  1--5.

\bibitem{Liu19}
J.~Liu, E.~Isufi, and G.~Leus, ``Filter design for autoregressive moving
  average graph filters,'' \emph{IEEE Trans. Signal, Info. Process. Netw.},
  vol.~5, no.~1, pp. 47--60, 2019.

\bibitem{Lu98}
W.-S. Lu, S.-C. Pei, and C.-C. Tseng, ``A weighted least-squares method for the
  design of stable {1-D and 2-D IIR} digital filters,'' \emph{IEEE Trans. Sig.
  Process.}, vol.~46, no.~1, pp. 1--10, Oct. 1998.

\bibitem{Shuman18}
D.~I. Shuman, P.~Vandergheynst, D.~Kressner, and P.~Frossard, ``Distributed
  signal processing via {Chebyshev} polynomial approximation,'' \emph{IEEE
  Trans. Signal, Info. Process. Netw.}, vol.~4, no.~4, pp. 736--751, 2018.

\bibitem{tseng21}
C.-C. Tseng and S.-L. Lee, ``Minimax design of graph filter using {Chebyshev}
  polynomial approximation,'' \emph{IEEE Trans. Circuits Syst. II, Exp.
  Briefs}, vol.~68, no.~5, pp. 1630--1634, May 2021.

\bibitem{Heath96ScientificCA}
M.~Heath and E.~Munson, \emph{Scientific Computing: An Introductory Survey},
  2nd~ed.\hskip 1em plus 0.5em minus 0.4em\relax Philadelphia: SIAM, 2018.

\bibitem{Lu2011a}
W.-S. Lu and T.~Hinamoto, ``Two-dimensional digital filters with sparse
  coefficients,'' \emph{Multidim. Syst. Signal Process.}, vol.~22, no. 1-3, pp.
  173--189, Mar. 2011.

\bibitem{Pak2021}
D.~Pakiyarajah, S.~S. Jayaweera, C.~U.~S. Edussooriya, C.~Wijenayake, and
  A.~Madanayake, ``{WLS} design of {M-D} complex-coefficient {FIR} filters with
  low group delay using second-order cone programming,'' in \emph{Proc. IEEE
  Int. Symp. Circuits Syst.}, 2021, pp. 1--5.

\bibitem{antoniou2007optimization}
A.~Antoniou and W.-S. Lu, \emph{Practical Optimization: Algorithms and
  Engineering Applications}.\hskip 1em plus 0.5em minus 0.4em\relax New York:
  Springer, 2007.

\bibitem{cvx}
M.~Grant and S.~Boyd, ``{CVX}: {Matlab} software for disciplined convex
  programming, version 2.1,'' \url{http://cvxr.com/cvx}, Mar. 2014.

\bibitem{gb08}
------, ``Graph implementations for nonsmooth convex programs.''\hskip 1em plus
  0.5em minus 0.4em\relax London: Springer-Verlag, 2008, pp. 95--110.

\bibitem{tseng20b}
C.-C. Tseng and S.-L. Lee, ``Design of graph filter using least-squares method
  with parameter norm penalty,'' in \emph{Proc. 2020 IEEE Int. Conf. Consumer
  Electronics - Taiwan (ICCE-Taiwan)}, 2020, pp. 1--2.

\end{thebibliography}
